\documentclass[preprint,12pt]{elsarticle}

\usepackage[cmex10]{amsmath}
\usepackage{amsfonts}
\usepackage{amssymb}
\usepackage{tikz}
\usepackage{algorithm,algorithmicx,algpseudocode}
\usepackage{subfigure}
\usepackage{url}
\def\N{\mathbb{N}}
\def\Z{\mathbb{Z}}

\def\B{\mathcal{B}}

\def\R{\mathbb{R}}

\def\card{\mathrm{card}}
\newproof{proof}{Proof}
\newtheorem{example}{Example}
\newtheorem{observation}{Observation}

\journal{}

\begin{document}

\begin{frontmatter}



\title{A local Gaussian filter and adaptive morphology as tools for completing partially discontinuous curves}


\author[label1]{P. Spurek}
\ead{przemyslaw.spurek@ii.uj.edu.pl}
\author[label1]{A. Chaikouskaya}
\author[label1]{J. Tabor}
\address[label1]{Faculty of Mathematics and Computer Science, 
Jagiellonian University, 
\L ojasiewicza 6, 
30-348 Krak\'ow, 
Poland}

\author[label2]{E. Zaj\c{a}c}
\address[label2]{Institute of Mathematics, 
Jan Kochanowski University,
\'Swi\c{e}tokrzyska 15, 
25-406 Kielce, 
Poland}
\begin{abstract}
This paper presents a method for extraction and analysis of curve--type structures which consist of disconnected components. 
Such structures are found in electron--microscopy (EM) images of metal nanograins, which are widely used in the field of nanosensor technology.

The topography of metal nanograins in compound nanomaterials is crucial to nanosensor characteristics. 
The method of completing such templates consists of three steps.
In the first step, a local Gaussian filter is used with different weights for each neighborhood. In the second step, an adaptive morphology operation is applied to detect the endpoints of curve segments and connect them.  
In the last step, pruning is employed to extract a curve which optimally fits the template.

\end{abstract}


\end{frontmatter}

\section{Introduction}

This paper presents a method for completing curve--type structures consisting of disjoint segments using Gaussian filter modification and adaptive morphology. Both of these methods modify images depending on the local properties. 

\begin{figure}[!t]
\centering
	\subfigure[]{\label{fig:ex_or_pic_1} 
\fbox{\includegraphics[width=2.55in]{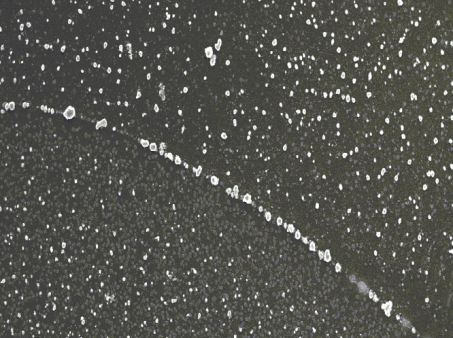}}} 
	\subfigure[]	
{\label{fig:ex_or_pic_2} 
\fbox{\includegraphics[width=2.3in]{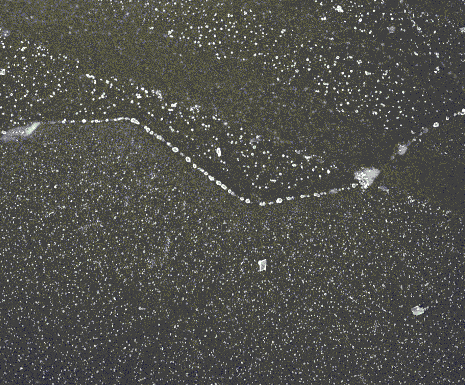}}} 
\caption{Original EM images of a carbonaceous--nanopalladium film.}
	\label{fig:ex_or_pic} 
\end{figure}

It is known that in some nanostructural films one can observe the formation of percolation paths \cite{n1}, which could be formed due to the influence of an electric or magnetic field on the film or, sometimes, due to film annealing. For example, such percolation paths have been observed for palladium-carbonaceous nanostructural films (nPd--C films) obtained by physical vapor deposition (PVD) in vacuum. The details of film preparation are described in paper \cite{n2}. The structure, morphology and topography of such films are described in papers \cite{n1,n3}. nPd--C films are composed of palladium nanograins (with a diameter of 1--4 nm), which are placed in a carbonaceous matrix. The effect of percolation paths generation has been observed for nPd--C films and is presented in paper \cite{n1}. These paths, composed of large agglomerated Pd grains, formed due to an electric field acting along the film surface. Such paths could be seen as wire-like structures, as it is shown in Fig. \ref{fig:ex_or_pic}. The bright objects visible in this image are palladium nanograins while the dark background is the carbonaceous matrix. The objective is to detect these curve-type (wire–type) templates in EM images. The investigated features consist of nanoparticle clusters (nanograins) with various sizes and geometries. Some of the nanograins are hidden below the surface of a thin layer (a carbonaceous matrix in Fig. \ref{fig:ex_or_pic}). Consequently, the wire--type nanostructures are represented in EM images as discontinuous curves with breaks of different width.

The proposed method consists of two steps: a local Gaussian filter and an adaptive morphological operation. 
The combination of these two methods allows one to fill in large gaps in curve templates. 
On the other hand, each of these methods gives good results in the case of a regular pattern with reasonably small breaks (see Figs. \ref{fig:road} and \ref{fig:dil_cov}). The basic difference between these methods is that the former uses grayscale images, while the latter works with black and white images. 

The first step is based on standard Gaussian filters \cite{RUSS,WF} and local scale selection \cite{Gomez}, which are used to deal with noise in signal and image processing \cite{Healey,Gerard,Berg,Geusebroek,Pavlovic,Kgg}. 
Our idea is based on blurring the elements of an image depending on the local properties, more precisely weighted covariance matrix.

\begin{figure}[!t]
\centering
	\subfigure[The result of a local Gaussian filter for the image presented in Fig. \ref{fig:ex_or_pic_1} with $r=15$.]{\label{fig:p2} 
\fbox{\includegraphics[width=1.5in]{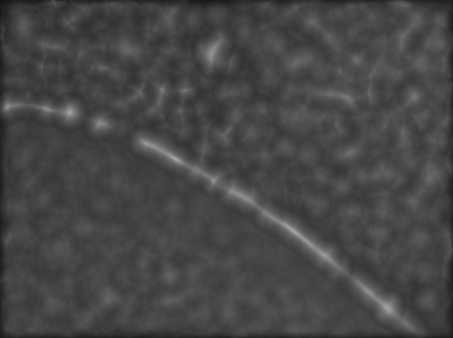}}} 
	\subfigure[After thresholding with a CEC algorithm.]{\label{fig:p3} 
\fbox{\includegraphics[width=1.5in]{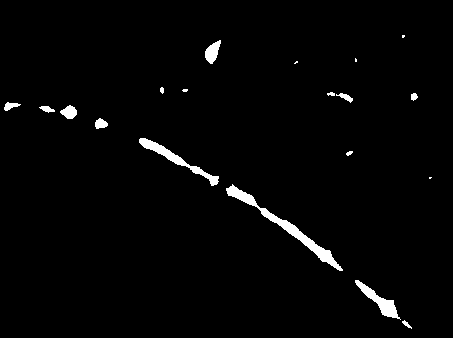}}}
	\subfigure[The result of an adaptive morphology filter with $r=15$ and $\alpha = 0.25$.]{\label{fig:p4} 
\fbox{\includegraphics[width=1.5in]{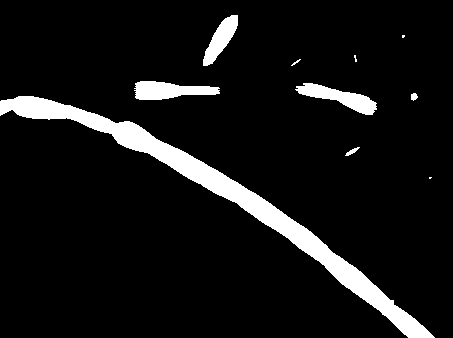}}}
	\subfigure[Skeletonization.]{\label{fig:p5} 
\fbox{\includegraphics[width=1.5in]{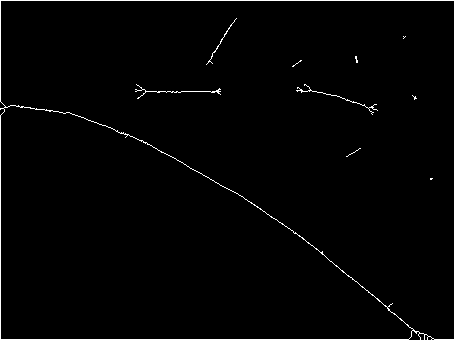}}}
	\subfigure[Pruning.]{\label{fig:p6} 
\fbox{\includegraphics[width=1.5in]{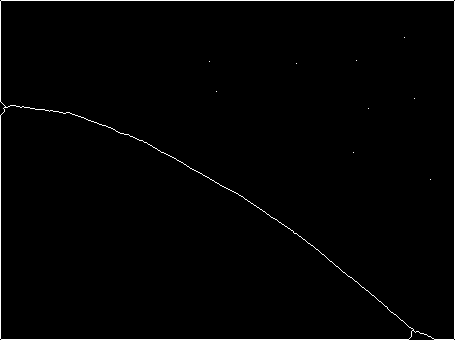}}}
\subfigure[The result of the proposed algorithm with segments connected to the final curve.]{\label{fig:p1} 
\fbox{\includegraphics[width=1.5in]{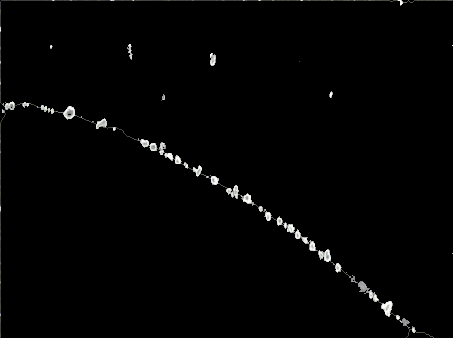}}} 
	\caption{Results of the proposed algorithm for the image presented in Fig. \ref{fig:ex_or_pic_1}.}
	\label{fig:ex_1} 
\end{figure}

The operation connects segments of a template and forms a blurred curve (see Fig. \ref{fig:p2}). On the other hand, the background, which consists of quite regular patterns, is transformed into an uniform layer.
Consequently, the curve template can be extracted by simple thresholding (see Fig. \ref{fig:p3}). 
In our paper CEC (cross entropy clustering) \cite{SpuTab,CEC1,CEC2,CEC3} algorithm is used for binarization.

In some cases, gaps are too large (see the left upper corner in Fig. \ref{fig:p3}) to be filled by a local Gaussian filter. To solve this problem, we used the other method -- a modification of classical mathematical morphology \cite{m_m_1,Shih,WR, mor_aply_1}.

This paper presents a new version of mathematical morphology which uses local covariance matrix. Similar approaches are presented in \cite{ad_mor_fill_2,ad_mor_fill_3,RUSS, mrze}, where segments are connected by locally estimated lines or curves. In \cite{ad_mor_fill_1}, the authors present a method based on a morphology operation which fills gaps by an elliptical structural element. Our approach also uses ellipses, but in contrast to the previous method, size and direction are extracted by using a covariance matrix instead of iteratively suited parameters. The results of our morphology are given in Fig. \ref{fig:p4}. 


The presented approach can also be used for different types of images. For instance, one can use it to connect parts of rivers and roads divided by bridges in satellite images of Earth  (see Fig. \ref{fig:road}) \cite{road_1,road_2}.

The next section of the paper presents a local Gaussian filter which uses different Gaussian functions in each point of an image. Then, an adaptive morphology operation is presented, which is applied to connect curve fragments. The last section presents the final version of our algorithm. 

\section{Local Gaussian filter}

Gaussian blur (also known as Gaussian smoothing) is the result of the blurring of an image by a Gaussian function. For processing images, one needs a two--dimensional Gaussian density distribution.
The normal random variable with the mean equal to zero
and the covariance matrix $\Sigma$ has a density
$$
g_{\Sigma}(x) := \frac{1}{2\pi \sqrt{ \det(\Sigma)}}\exp(-\frac{1}{2}\| x \|_{\Sigma}^2)
$$
where by  
$\|x\|_{\Sigma}^2$ we denote the Mahalanobis norm
\cite{Ma} of $x \in \R^2$ 
$$
\| x \|_{\Sigma}^2:=x^T\Sigma^{-1}x.
$$

\begin{figure}[!t]
\centering
	\subfigure[Neighborhood for $r = 1$.]{\label{fig:s_1} 
\begin{tikzpicture}[scale=4]

\draw [fill=lightgray] (0.0,0.1) rectangle (-0.1,-0.0);

\draw[line width=1pt] (-0.2,0.1) -- (-0.2,-0.0);
\draw[line width=1pt] (-0.1,0.2) -- (0.0,0.2);

\draw[line width=1pt] (-0.1,-0.1) -- (0.0,-0.1);
\draw[line width=1pt] (-0.1,-0.1) -- (-0.1,0.2);
\draw[line width=1pt] (-0.2,0.1) -- (0.1,0.1);
\draw[line width=1pt] (0.1,0.1) -- (0.1,-0.0);

\draw[line width=1pt] (-0.2,0.0) -- (0.1,0.0);
\draw[line width=1pt] (0.0,0.2) -- (0.0,-0.1);

\end{tikzpicture}}
\qquad \qquad
	\subfigure[Neighborhood for $r = 2$.]{\label{fig:s_2} 
\begin{tikzpicture}[scale=4]

\draw [fill=lightgray] (0.0,0.1) rectangle (-0.1,-0.0);

\draw[line width=1pt] (-0.3,-0.1) -- (-0.3,0.2);
\draw[line width=1pt] (-0.2,0.3) -- (0.1,0.3);

\draw[line width=1pt] (-0.2,-0.2) -- (0.1,-0.2);
\draw[line width=1pt] (-0.2,0.3) -- (-0.2,-0.2);
\draw[line width=1pt] (-0.3,0.2) -- (0.2,0.2);
\draw[line width=1pt] (0.2,0.2) -- (0.2,-0.1);

\draw[line width=1pt] (-0.3,-0.1) -- (0.2,-0.1);
\draw[line width=1pt] (-0.1,-0.2) -- (-0.1,0.3);
\draw[line width=1pt] (-0.3,0.1) -- (0.2,0.1);
\draw[line width=1pt] (0.1,0.3) -- (0.1,-0.2);

\draw[line width=1pt] (-0.3,0.0) -- (0.2,0.0);
\draw[line width=1pt] (0.0,0.3) -- (0.0,-0.2);

\end{tikzpicture}}
\qquad \qquad
	\subfigure[Neighborhood for $r = 3$.]{\label{fig:s_3} 
\begin{tikzpicture}[scale=4]

\draw [fill=lightgray] (0.0,0.1) rectangle (-0.1,-0.0);

\draw[line width=1pt] (-0.3,-0.2) -- (-0.3,0.3);
\draw[line width=1pt] (-0.3,0.3) -- (0.2,0.3);

\draw[line width=1pt] (-0.3,-0.2) -- (0.2,-0.2);
\draw[line width=1pt] (-0.2,0.4) -- (-0.2,-0.3);
\draw[line width=1pt] (-0.4,0.2) -- (0.3,0.2);
\draw[line width=1pt] (0.2,0.3) -- (0.2,-0.2);

\draw[line width=1pt] (0.3,-0.1) -- (0.3,0.2);
\draw[line width=1pt] (-0.4,-0.1) -- (-0.4,0.2);

\draw[line width=1pt] (-0.4,-0.1) -- (0.3,-0.1);
\draw[line width=1pt] (-0.1,-0.3) -- (-0.1,0.4);
\draw[line width=1pt] (-0.4,0.1) -- (0.3,0.1);
\draw[line width=1pt] (0.1,0.4) -- (0.1,-0.3);

\draw[line width=1pt] (-0.4,0.0) -- (0.3,0.0);
\draw[line width=1pt] (0.0,0.4) -- (0.0,-0.3);

\draw[line width=1pt] (-0.2,-0.3) -- (0.1,-0.3);
\draw[line width=1pt] (-0.2,0.4) -- (0.1,0.4);

\end{tikzpicture}}
\caption{Neighborhoods with different $r$.}
\label{fig:morph_s}
\end{figure}
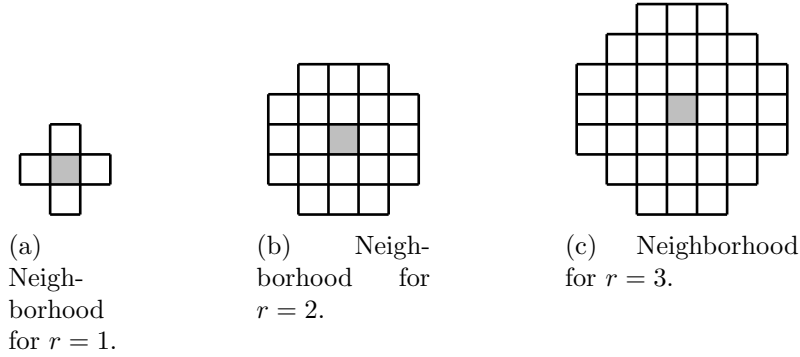

An image with dimensions $m \times n$ is interpreted as a function $J \colon [0,m-1] \times [0,n-1] \to [0,1]$, where $J(k_1,k_2 )$ describes the intensity of the pixel with coordinates $(k_1,k_2 )$. 
In this paper, the function is extended to $J \colon \Z \times \Z \to [0,1]$ by assigning $J(k_1,k_2 ) = 0$ for $(k_1,k_2) \notin [0,m-1] \times [0,n-1]$.

All calculations are dedicated to the circular neighborhood
 (see Fig. \ref{fig:morph_s}) of a fixed point $(k_1,k_2) \in \Z \times \Z$. A circle with its center at zero  and a radius $r>0$ is denoted by 
$$
\B_r : = \{ (i,j) \in \Z^2 \colon i^2+j^2 \leq r^2  \}.
$$

Classical Gaussian blur is based on a convolution operation with a mask of the size $2 \cdot r+1$ for $r \in \N$.  Consequently, each pixel with coordinates $(k_1,k_2) \in \Z \times \Z$ is transformed into

$$
( g_{\Sigma} \ * \ J ) (k_1,k_2) : = \sum_{\substack{i,j=-r \\ i^2+j^2 \leq r^2 }}^{r}  g_{\Sigma}(i,j) J(k_1+i, k_2 + j )
$$
where $\Sigma$ is fixed.

\begin{figure}[!t]
\centering
	\subfigure[Gaussian blur with $s=1$.]{\label{fig:blu_clasicr_1} 
\fbox{\includegraphics[width=1.6in]{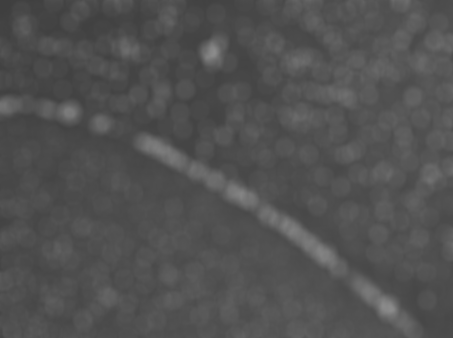}}}
	\subfigure[Otsu thresholding of Fig \ref{fig:blu_clasicr_1}.]{\label{fig:blu_clasicr_2} 
\fbox{\includegraphics[width=1.6in]{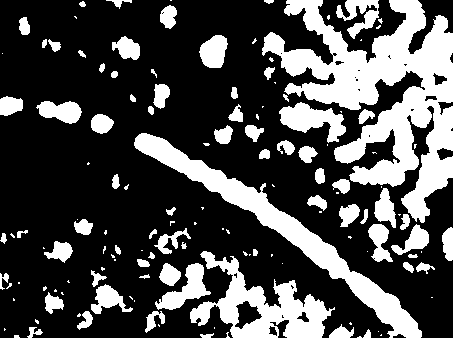}}} 
	\subfigure[CEC thresholding of Fig \ref{fig:blu_clasicr_1}.]{\label{fig:blu_clasicr_3} 
\fbox{\includegraphics[width=1.6in]{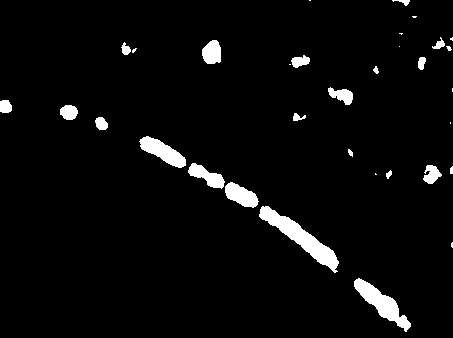}}} 
	\caption{Gaussian blur with $2r+1$ windows and covariance matrix $sI$.}
	\label{fig:blur_clasic} 
\end{figure}

\begin{figure}[!t]
\centering
	\subfigure[An example of an EM image.]{\label{fig:blur_1} 
\fbox{\includegraphics[width=2.2in]{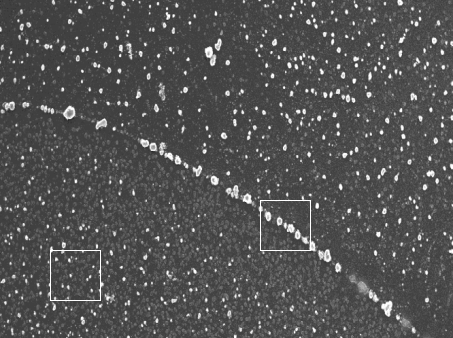}}}
\qquad
	\subfigure[The result of a local Gaussian filter.]{\label{fig:blur_2} 
\fbox{\includegraphics[width=2.2in]{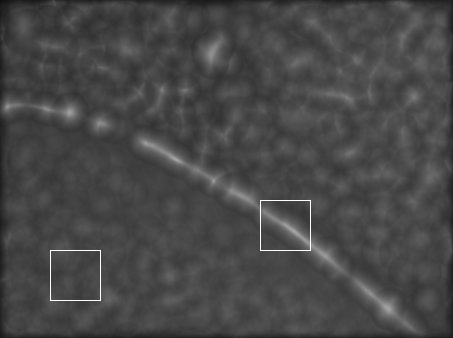}}} \\
	\subfigure[Levels for the Gaussian function estimated for the detail marked on the left.]{\label{fig:blur_3}
\fbox{\includegraphics[width=1.6in]{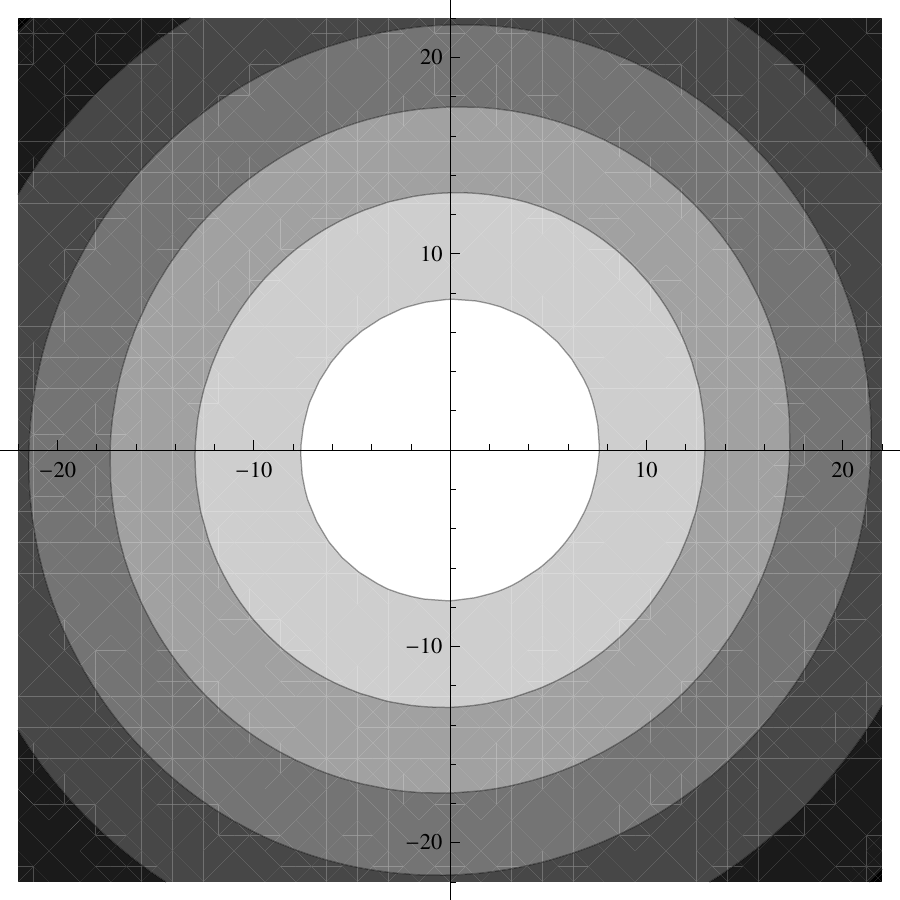}}}
\qquad
	\subfigure[Levels for the Gaussian function estimated for the detail marked on the right.]{\label{fig:blur_4}
\fbox{\includegraphics[width=1.6in]{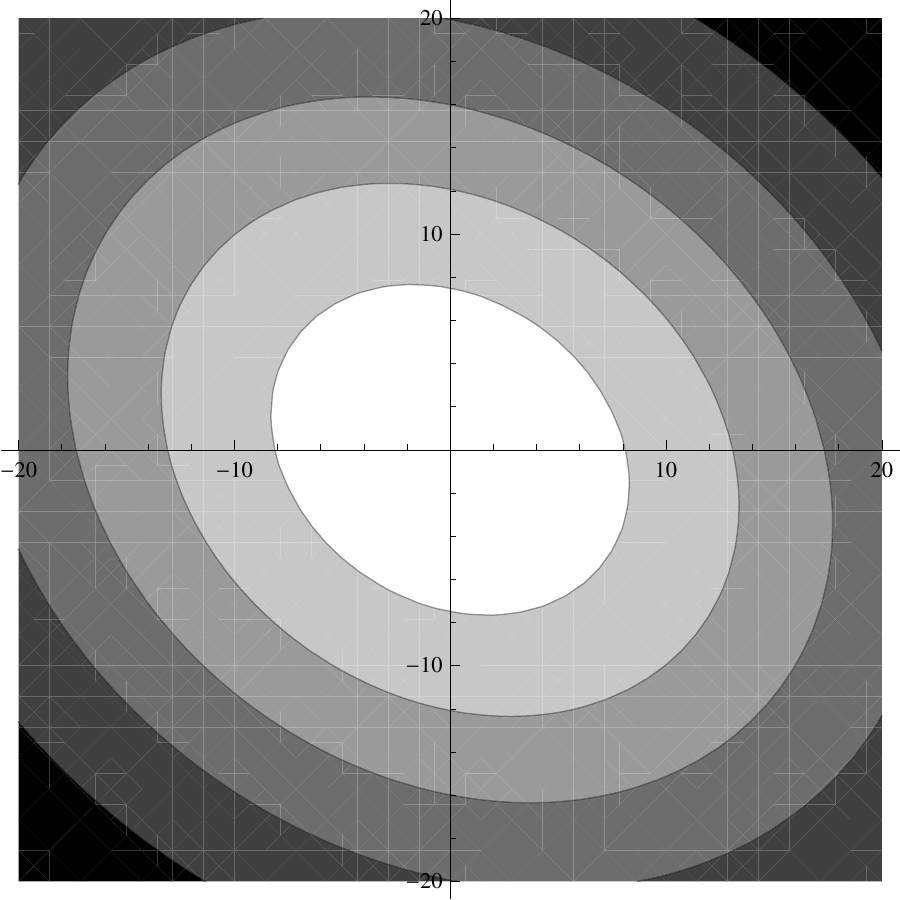}}}
	\caption{The results of the proposed algorithm.}
	\label{fig:blur} 
\end{figure}

In the classical approach, $\Sigma$ is proportional to the identity matrix. 
The results of the standard Gaussian filter in the case of the image from Fig. \ref{fig:ex_or_pic_1} are presented in Fig. \ref{fig:blur_clasic}. As it can be seen, when one uses a covariance matrix proportional to the identity matrix, the curve template is uniformly blurred in all directions (see Fig. \ref{fig:blu_clasicr_2}). Moreover, one obtains more holes which are difficult to reduce (compare Fig. \ref{fig:p3} and Fig. \ref{fig:blu_clasicr_3}).

As it was said, in this paper CEC thresholding \cite{SpuTab,CEC1} is used instead of the classical Otsu  \cite{otsu} method. Fig. \ref{fig:blur_clasic} presents the basic differences between these two approaches (compare Fig. \ref{fig:blu_clasicr_2}) and  Fig. \ref{fig:blu_clasicr_3}).

\begin{figure}[!t]
\centering
\fbox{\includegraphics[width=1.2in]{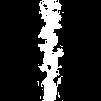}}
\fbox{\includegraphics[width=1.2in]{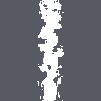}}
\fbox{\includegraphics[width=1.2in]{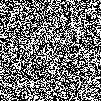}}\\
\fbox{\includegraphics[width=1.2in]{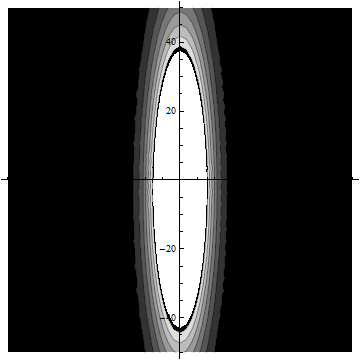}}
\fbox{\includegraphics[width=1.2in]{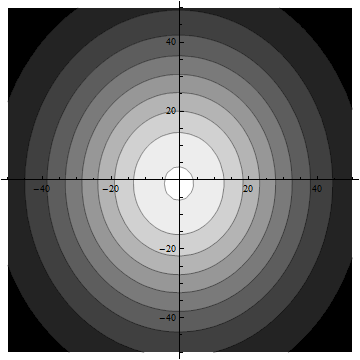}}
\fbox{\includegraphics[width=1.2in]{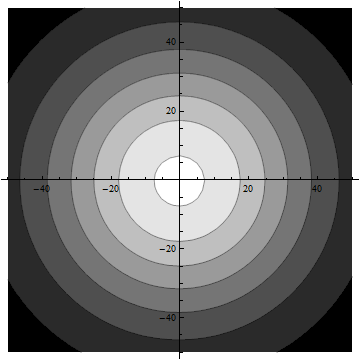}}
\caption{Examples of simple images and Gaussian functions.}
\label{fig:cov_temp}
\end{figure}

Our method uses the coordinates of pixels and grayscale color. More precisely, the intensities of pixels in a neighborhood are used as weights. Note that a small change of background intensity radically deforms the shape of the estimated Gaussian function (see Fig. \ref{fig:cov_temp}). 

Since the color range of the image has major influence on the shape of the estimated Gaussian function, we reduce it by subtracting the mean color in a neighborhood and taking the maximum from point zero and the mean. 
Consequently, in the neighborhood of point $(k_1,k_2)$ the following function is considered:
$$
J_{k_1,k_2}^{r} (l,k)  :=  \max \left\{ 0, J(l,k)   -   \frac{ 1 }{\card(\B_r)} \sum \limits_{i,j \in \B_r }  J(k_1   +  i, k_2  +  j )  \right\}.
$$

In the case of EM images of a carbonaceous-nanopalladium film, one often deals with curves which separate two environments with different particle distributions (see Fig. \ref{fig:p3}). This effect causes a distortion of the covariance matrix. Moreover, the average of pixels (from the circular neighborhood) is far from the center of neighborhood. Consequently, it is impossible to detect the main direction of the curve. To deal with these problems, we consider not only points from the circular neighborhood $\B_r$, but also their inversion through the center. For a fixed point $(k_1,k_2)$ and an element in the neighborhood $(i,j)$, we use the original point with color $J_{k_1,k_2}^r(k_1+j,k_2+j)$ and a symmetrical one  $(-i,-j)$ with the same weight $J_{k_1,k_2}^r(k_1+j,k_2+j)$.

Therefore, in the case of the circular neighborhood $\B_r$ at point  $(k_1,k_2) \in \Z \times \Z$, a weighted covariance matrix is given by
\begin{equation}\label{equ:cov}
 \Sigma_{J}^{r}(k_1,k_2)  = 
    \frac{\sum \limits_{i,j \in \B_r }   J_{k_1,k_2}^{r} (k_1+i,k_2+j) \cdot [(i, j)^{T}  \cdot  ( i, j)] }{  \sum \limits_{i,j \in \B_r }    J_{k_1,k_2}^{r}(k_1+i,k_2+j)   } 
\end{equation}

Gaussian blur is obtained by replacing each pixel with coordinates $(k_1,k_2)$ using the formula 
$$
\left( g_{\Sigma} * J \right) (k_1,k_2),
$$
where $\Sigma = \Sigma_{J}^{r}(k_1,k_2)$.

The effects of the proposed filter for EM images are presented in Fig. \ref{fig:blur}. This figure also shows differences between Gaussian density estimated from an element of the image containing background (see Fig. \ref{fig:blur_3}) and part of the curve--type structure (see Fig. \ref{fig:blur_4}). 

Description of our algorithm is presented in Algorithm \ref{alg1}.

\begin{algorithm} 
\caption{Local Gaussian blur:}          
\label{alg1}    
		\begin{algorithmic}
			\State {\bf input}
			\State \textit{$J \colon [0,m-1] \times [0,n-1] \to [0,1]$} \Comment{input image} 	
			\State \textit{ $I \colon [0,m-1] \times [0,n-1] \to [0,1]$}	 \Comment{output image}		
			\State \textit{ $r > 0$ } \Comment{radius of the structural element}			
			\For{$ (k_1,k_2) \in [0,m-1] \times [0,n-1] $}
				\State \textit{$\Sigma \gets \Sigma_{J}^{r}(k_1,k_2)$ } \Comment{calculate local covariance} 
				\State $I(k_1,k_2) \gets ( g_{\Sigma} * J ) (k_1,k_2)$ 
 			\EndFor
		\end{algorithmic}
\end{algorithm}

%
%

\begin{figure}[!t]
\centering
	\subfigure[Original image.]{\label{fig:road_1} 
\fbox{\includegraphics[width=1.6in]{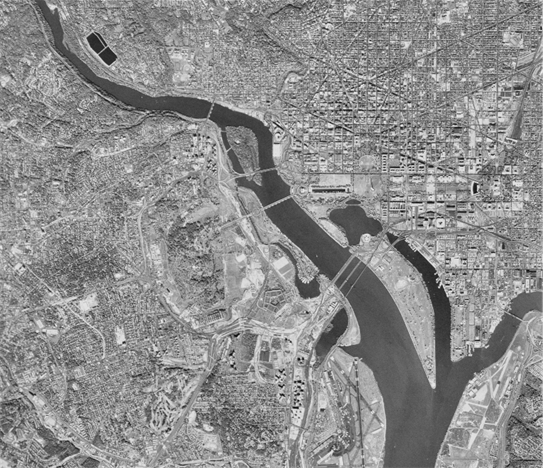}}}
	\subfigure[Local Gaussian filter with the neighborhood $r = 21$.]{\label{fig:road_3}
\fbox{\includegraphics[width=1.6in]{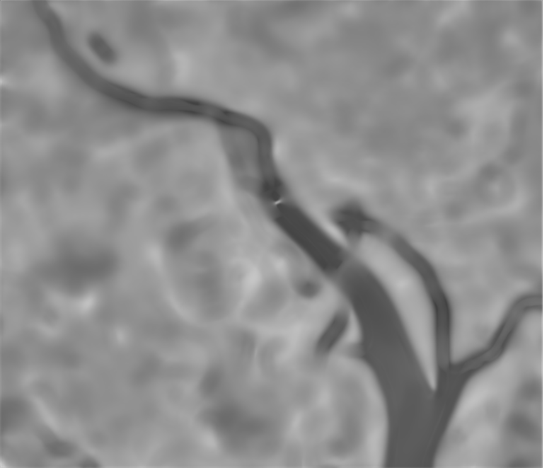}}}\\
	\subfigure[Thresholding with a CEC algorithm.]{\label{fig:road_4}
\fbox{\includegraphics[width=1.6in]{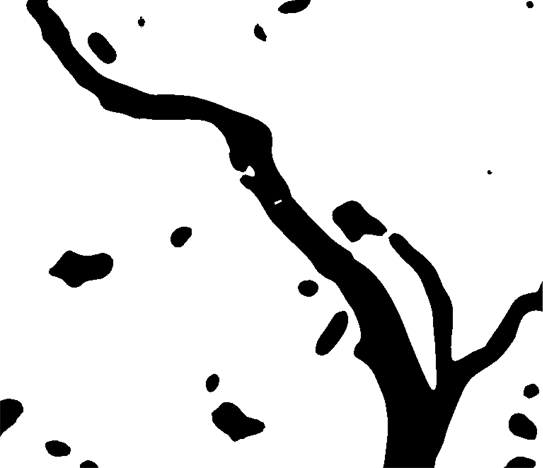}}}
	\subfigure[Original image and the result of the proposed algorithm (after skeletonization and
pruning).]{\label{fig:road_4}
\fbox{\includegraphics[width=1.6in]{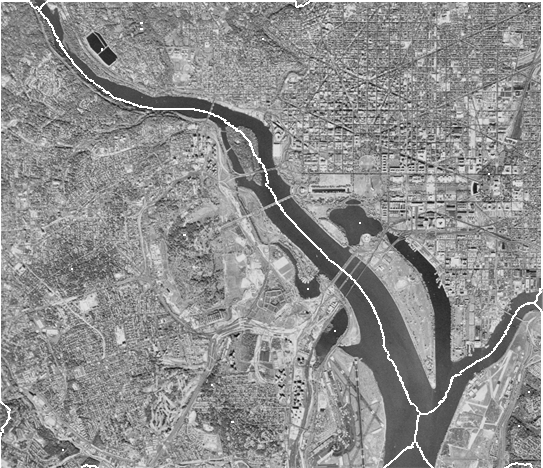}}}

\caption{Line detection.}
\label{fig:road}
\end{figure}

 At the end of this section, a possible application of local Gaussian blur is presented.
As was said in the introduction, the proposed algorithm can be used for locating curve structures such as rivers and roads, in satellite images (see Fig. \ref{fig:road}). Our method identifies the main part of the river and can deal with barriers such as bridges and roads. In future, our method can be used to estimate the length of rivers. 

In the case of templates containing equal elements with similar gaps our method gives good results. Unfortunately, EM images of carbonaceous--nanopalladium films present a more complicated situation. Curves contain segments of different size and geometry. Moreover, the gaps which need to be filled have different widths. 

A local Gaussian filter and thresholding led to the elimination of most gaps, but some still persisted in the studied curve--type structure. An adaptive morphology operation will be used to deal with that problem. 

\section{Adaptive morphology operation}

\begin{figure}[!t]
\centering
	\subfigure[Classical dilation for the image from Fig. \ref{fig:p3}. Minimal possible windows  ($r=31$) ) are used to fill the gaps in the curve.]{\label{fig:line_2_bin_di} 
\fbox{\includegraphics[width=2.0in]{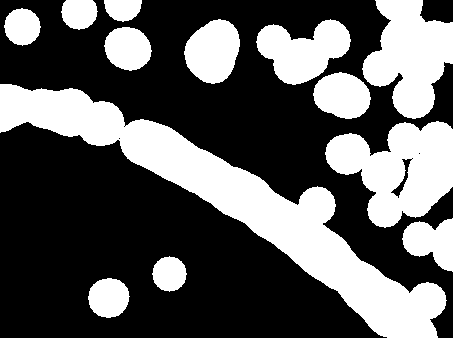}}}
\qquad
	\subfigure[Adaptive local dilation for the image from Fig. \ref{fig:p3} with $r=25$ and $\alpha = 0.25$.] 
{\label{fig:line_3_bin_di} 
\fbox{\includegraphics[width=2.0in]{example_n/15_CEC_25_0_25.png}}}
	\caption{Results of the classical version and our modification of morphology operations for the image from Fig. \ref{fig:p3}.}
	\label{fig:line_bin_di} 
\end{figure}

In the case of EM images, curves cannot be detected directly and some gaps remain after thresholding (see Fig. \ref{fig:p3}). To solve this problem, a mathematical morphology operation is usually applied \cite{m_m_1,Shih,WR}, which uses a fixed structural element in all pixels, although the size and shape of that element can be arbitrarily designed. Unfortunately, classical dilation increases data in all directions (see Fig. \ref{fig:line_2_bin_di}).

This section presents an adaptive version of morphology and its application for filling gaps between segments (see Fig. \ref{fig:line_3_bin_di}). Similarly as in the local Gaussian filter, local properties of images are used. More precisely, a covariance matrix is employed to fit the size and orientation of elliptical structural elements. Since the set should be extended only in the direction of the gaps, a morphology operation is used only if necessary. 

Consequently, for each point one needs to answer the questions: {\em Should the morphology operation be applied in this place?} and {\em What size and orientation of the elliptical structural element should be used?}
 
In this section, similar to the previous one, a covariance matrix is used to address these questions. Before proceeding, let us recall some basic information about 2--dimensional ellipses.

Let $\Sigma$ be a positive definite matrix of size $2 \times 2$. The $2$--dimensional ellipse generated by matrix $\Sigma$ with its center in zero is defined as follows 
$$
\B_{\Sigma} := \left\{ (k_1,k_2) \in \Z^2 \colon \| (k_1,k_2) \|_{\Sigma}^2 < 1  \right\}.
$$ 
The eigenvectors of $\Sigma$ define the principal directions of the ellipse 
and the eigenvalues of $\Sigma$ are the squares of the semi--axes: ${a^2}$, ${b^2}$. 
On the other hand, the covariance matrix of uniform density of an ellipse
$$
 \left\{ (k_1,k_2) \in \Z^2 \colon \frac{k_1^2}{a^2} + \frac{k_2^2}{b^2} < 1  \right\}
$$
is given by 
$$
\Sigma_{a,b} := \left[ \begin{array}{cc}
\frac{a^2}{4} & 0 \\
0 & \frac{b^2}{4} \\
\end{array} \right]. 
$$
Therefore, elliptical structural elements with radiuses 
$2 \sqrt{\lambda_1}$, $2 \sqrt{\lambda_2}$ and the principal directions $v_1$, $v_2$ are used, where $\lambda_1$, $\lambda_2$ are eigenvalues and $v_1$, $v_2$ are eigenvectors of the covariance matrix. In other words, the square roots of eigenvectors are proportional to parameters $a,b$:
$$
\frac{\sqrt{\lambda_2}}{\sqrt{\lambda_1}} = \frac{a}{b}, \mbox{ assuming } a \leq b.
$$

The following observation presents a method for extracting an ellipse without eigendecomposition of $\Sigma$.
\begin{observation}
Let $a,b \in \R$ be given, then
$$
\left\{ (k_1,k_2) \colon \frac{k_1^2}{a^2} + \frac{k_2^2}{b^2} < 1 \right\} = \left\{ (k_1,k_2) \colon \|(k_1,k_2)  \|_{ \Sigma_{a,b} }^2 < 4 \right\}
$$
\end{observation}

\begin{proof}
By simple calculations, one obtains 
$$
\begin{array} {l}
\|(k_1,k_2)  \|_{ \Sigma_{a,b} }^2 =
(k_1,k_2)^T\Sigma_{a,b}^{-1}(k_1,k_2) =  
 \hspace{2ex} \\
(k_1,k_2)^T \left[ \begin{array}{cc}
\frac{4}{a^2} & 0 \\
0 & \frac{4}{b^2} \\
\end{array} \right]
(k_1,k_2) = 
\frac{4k_1^2}{a^2} + \frac{4k_2^2}{b^2}. 
\end{array}
$$
\end{proof}

Thanks to these observations, one can easily draw an ellipse using only knowledge about $\Sigma$. More precisely, for point $(k_1,k_2)$ and a fixed $r$, 
the structural element is given by the following formula:
$$
\{ (k_1,k_2)\in \Z^2 \colon  \| (k_1+i,k_2+j) \|_{\Sigma_{J}^{r}(k_1,k_2)}^2 < 4 \}.
$$

Now let us return to the proposed algorithm. Again, a neighborhood of size $2r+1$ is used and the covariance matrix is determined by the formula (\ref{equ:cov}). 

\begin{figure}[!t]
\begin{center}
\input{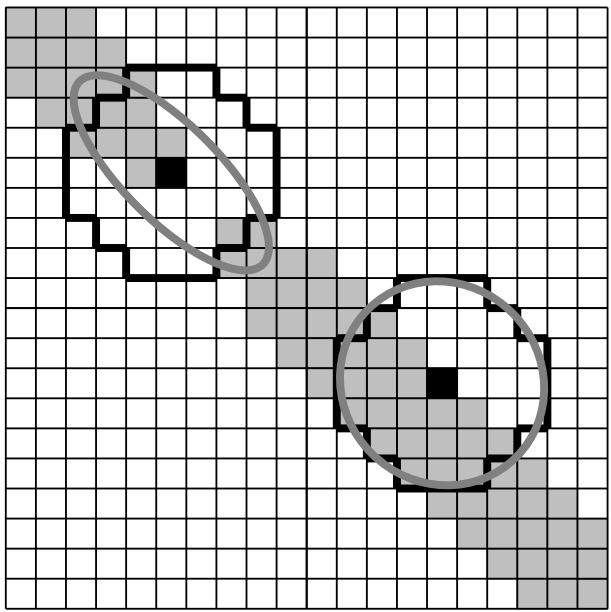}
\end{center}
\caption{Structural elements adjusted to different situations.}
\label{fig:morph}
\end{figure}

Since the set should be increased only in the direction of the gap, one needs to verify whether the pixels in question are endpoints of a segment or intermediary points. This is explained in greater detail in the example: 

\begin{example}\label{ex:mor_1}
Let us consider the set presented in Fig.~\ref{fig:morph}. Let $r=2$ be arbitrarily fixed. The behavior of an adaptive morphology is described for two points, which are marked black. 

We use coordinates of points from the neighborhood to determine a covariance matrix. In this example, elements of the curve are marked in light gray. 

In the case of the first point (upper left corner in Fig. \ref{fig:morph})
the 
 eigenvalues of the covariance matrix are: $\lambda_1 = 4.6$ and $\lambda_2 = 0.6$. 
Moreover, $\frac{ \sqrt{ \lambda_2 } }{ \sqrt{ \lambda_1} } = 0.36 $, so the structural element in this point is elliptical. Thus, the gap is filled without expanding our set in all directions.

In the second case, 
the covariance matrix is 
$ 
\left[ \begin{array}{cc}
2.9 & 0.1 \\
0.1 & 2.9 \\
\end{array} \right]. 
$
The eigenvalues are: $\lambda_1 = 3$ and $\lambda_2 = 2.7$. Consequently, we obtain $\frac{ \sqrt{ \lambda_2 } }{ \sqrt{ \lambda_1} } = 0.95$ which means that the structural element is circular.
In general, circular structural elements expand the curve in all direction. Consequently, we do not want to apply morphology operation in this place.

%
\end{example}


Thus, a morphological operation is used only if eigenvalues are clearly different, or,
more precisely, if $\frac{ \sqrt{ \lambda_2}}{ \sqrt{\lambda_1}} \leq \alpha $ where $\alpha$ is fixed parameter. 

The value of parameter $\alpha$ has a considerable influence on the final result of the morphology operation.
Unfortunately, it is difficult to choose an optimal value.
In fact, it depends on the size of the neighborhood and the geometry of the largest gaps that need to be filled. 

\begin{figure}[!t]
\centering
\fbox{\includegraphics[width=2.5in]{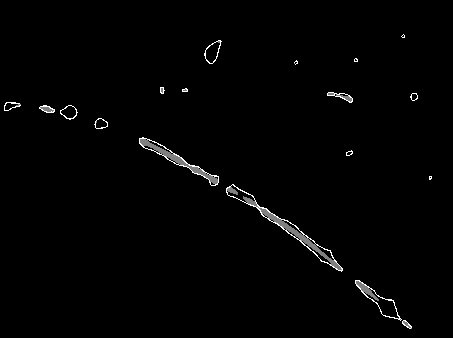}}
\caption{Contours of elements from Fig. \ref{fig:p3}. ). Points in which a morphology operation was applied are marked in gray.}
\label{fig:condi}
\end{figure}

Fig. \ref{fig:condi} presents points in which a morphology operation is applied, for the image from Fig. \ref{fig:p3}.

\begin{figure}[!t]
\centering
	\subfigure[Original image.]{\label{fig:dil} 
\fbox{\includegraphics[width=1.2in]{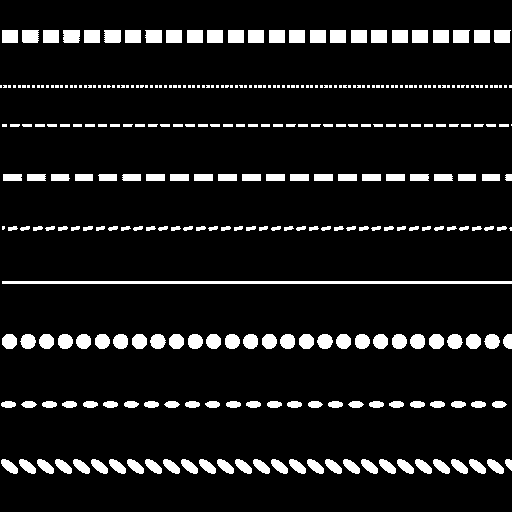}}}
	\subfigure[Local adaptive dilation for a neighborhood with $r = 25$ and $\alpha = 0.3$.]	{\label{fig:dil_1}
\fbox{\includegraphics[width=1.2in]{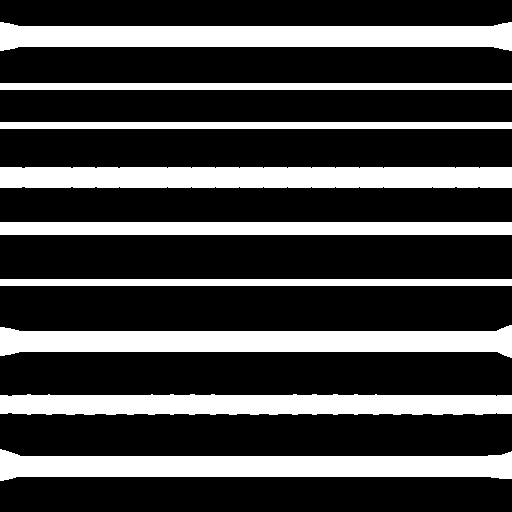}}}
	\subfigure[Local adaptive dilation for a neighborhood with  $r = 25$ and $\alpha = 1$.]{\label{fig:dil_2}
\fbox{\includegraphics[width=1.2in]{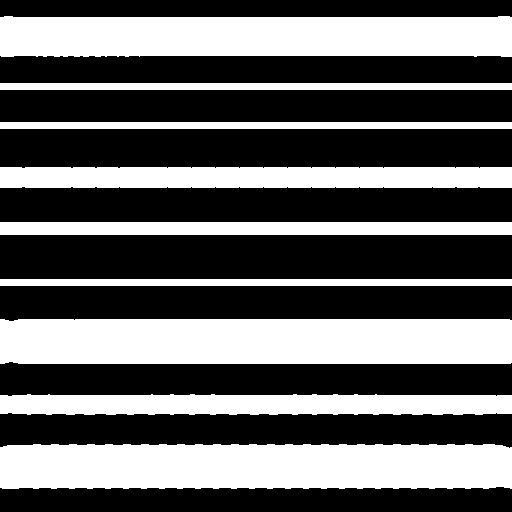}}}
	\subfigure[Classical dilation for a neighborhood with $r = 21$.]{\label{fig:dil_3}
\fbox{\includegraphics[width=1.2in]{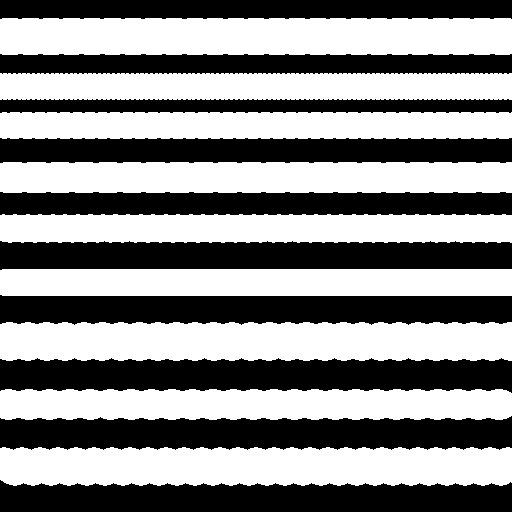}}}
\caption{Line detection.}
\label{fig:dil_cov}
\end{figure}

Let us consider one more example. Fig. \ref{fig:dil} shows different types of lines. The results of the proposed adaptive morphology operation are presented in Fig. \ref{fig:dil_1} and Fig. \ref{fig:dil_2}, while the results of classical morphology operations are given in Fig. \ref{fig:dil_3}. The adaptive version allows one to increase the set in the direction of the gap, while the standard approach expands the set in all directions. 

\begin{algorithm} 
\caption{Adaptive morphology operation:}          
\label{alg2}    
		\begin{algorithmic}
			\State {\bf input}
			\State \textit{$J \colon [0,m-1] \times [0,n-1] \to [0,1]$} \Comment{black and white image} 		
			\State \textit{$I \colon [0,m-1] \times [0,n-1] \to [0,1]$}	 \Comment{output, copy of $J$}	
			\State \textit{ $r > 0$ } \Comment{radius of the structural element}			
			\State \textit{ $\alpha \in [0,1]$ } 
			\For{$ (k_1,k_2) \in [0,m-1] \times [0,n-1]$}
\If{$J(k_1,k_2) = 1$} \Comment{only in white pixels}
				\State \textit{$\Sigma \gets \Sigma_{J}^{r}(k_1,k_2)$ } \Comment{calculate local covariance} 
				\State \textit{$\lambda_1, \lambda_2 \gets \mathrm{Eig}( \Sigma)$ } \Comment{eigenvalues ($\lambda_1 \geq \lambda_2$)} 
\If{$\frac{\lambda_2}{\lambda_1} \leq \alpha$}
				\For{$ (i,j) \in [-r,r] \times [-r,r] $}
			     \If {$ \|(i,j)  \|_{ \Sigma }^2 < 4 $}
				\State \textit{$I(k_1+i,k_2+j) \gets 1$}
				\EndIf
 			     \EndFor
	
\EndIf				
\EndIf
 			\EndFor
		\end{algorithmic}
\end{algorithm} 

Algorithm 2 presents the pseudo code of the adaptive morphology method.

\section{The final method}

The main objective of the presented algorithm is to detect curve-type structures in EM images of compound nanomaterials, such as carbonaceous--nanopalladium films. In the case of this kind of images one deals with small elements of different shapes and geometries which are separated by gaps of different sizes.

The presented method consists of two steps. In general, in simple situations (see Fig. \ref{fig:road}) it is sufficient to use one of them. Local Gaussian blur uses the grayscale color of images, since an adaptive morphology operation requires a binary version of images. 
 
Since the procedure should lead to producing a curve, skeletonization is applied \cite{skel1}. As after local morphology operations images exhibit some connected components with non-standard shapes, we use pruning \cite{pruning1,pruning2} and thinning \cite{thin1}.

Our method can be described as follows:
\begin{enumerate}
\item apply a local Gaussian filter,
\item apply CEC thresholding,
\item apply an adaptive morphology filter, 
\item apply skeletonization, thinning and pruning to extract a curve which describes the studied curve--type nanograin structures.
\end{enumerate}

The presented algorithm requires three parameters. First of all, one needs to define the size of neighborhood for Gaussian blur and adaptive morphology operation. The same value can be used for both methods. This parameter should be approximately equal to the largest gap in the structure in question. On the other hand, one needs parameter $\alpha$ for morphology operations. 

\begin{figure}[!t]
\centering
	\subfigure[The result of local Gaussian blur with $r=10$ for the image presented in Fig.  \ref{fig:ex_or_pic_2}.]
{\label{fig:e_3_15} 
\fbox{\includegraphics[width=1.6in]{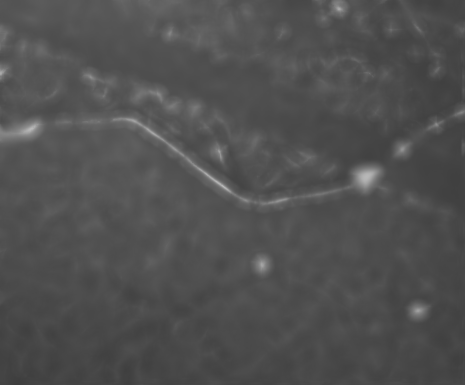}}}
	\subfigure[After thresholding with a CEC algorithm.]
{\label{fig:e_3_15_90} 
\fbox{\includegraphics[width=1.6in]{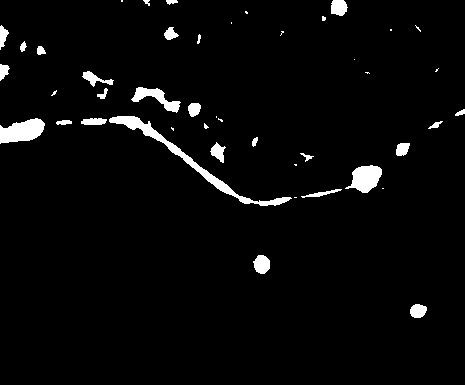}}}
	\subfigure[The result of an adaptive morphology filter with $r=25$ and $\alpha = 0.3$.]
{\label{fig:e_3_15_skel} 
\fbox{\includegraphics[width=1.6in]{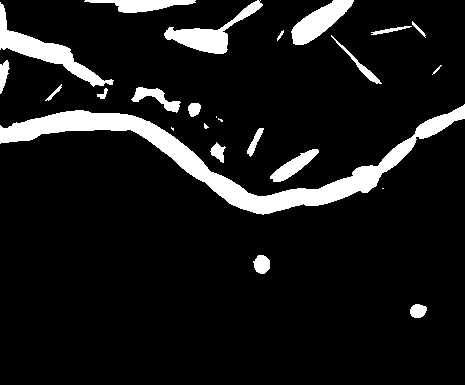}}}
	\subfigure[Skeletonization.]
{\label{fig:e_3_15_skel} 
\fbox{\includegraphics[width=1.6in]{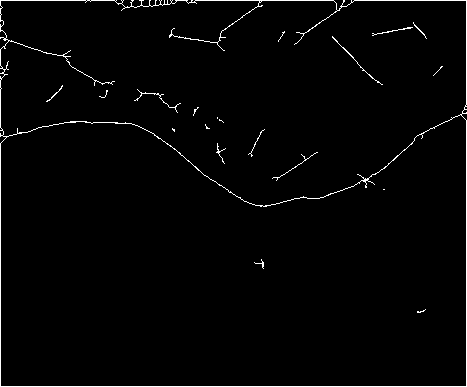}}}
	\subfigure[Pruning.]
{\label{fig:e_3_15_skel} 
\fbox{\includegraphics[width=1.6in]{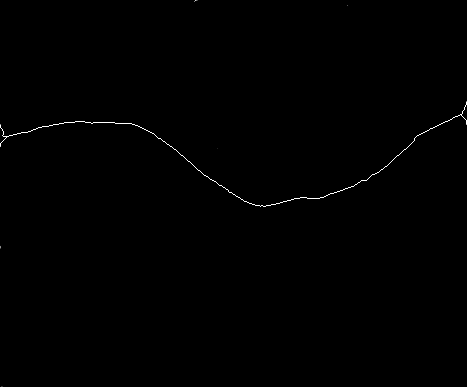}}}
	\subfigure[The result of the proposed algorithm with segments connected to the final curve.]
{\label{fig:e_3_ro} 
\fbox{\includegraphics[width=1.6in]{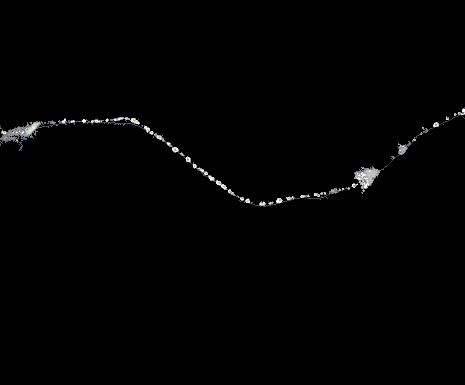}}}

\caption{Examples of detecting a curve--type template for the image presented in Fig. \ref{fig:ex_or_pic_2}.}
\label{fig:e_3}
\end{figure}

The results of our method for EM images are presented in Fig. \ref{fig:e_3}. 

\section{Conclusion}

This paper presents a method for completing curve--type structures consisting of small elements of different shapes and geometries. In general, two approaches to this problem are presented. One is based on local Gaussian blur. The use of weighted covariance makes it possible to find the direction of the position of the next element locally. The objective is to blur curve segments so that they could be connected.

The other method is based on an adaptive morphology operation with elliptical structural elements whose size and orientation are locally estimated by the use of a covariance matrix. Moreover, thanks to spectral analysis of the covariance matrix, one can verify which pixels are endpoints of a given curve element (dilation is applied) or intermediate points (a morphological operation is not applied).
 
By using a combination of these two methods, it is possible to detect curves, i.e., wire-type nanograin structures in EM images of compound nanomaterials. Furthermore, the presented method can be applied in various situations, such as road and river detection in satellite images. 

Implementation of local Gaussian filter and Adaptive morphology operation as a plug--in for imageJ is available in \cite{imageJ}.

\section{Acknowledgments}

This work was partially (the part of E. Zaj¹c) supported by the European Regional Development Fund within the 2007–2013 Innovative Economy Operational Programme (the project "Development of technology for a new generation of the hydrogen and hydrogen compounds sensor for applications in above normative conditions," No UDA-POIG.01.03.01-14-071/08-09).












\begin{thebibliography}{33}
\expandafter\ifx\csname natexlab\endcsname\relax\def\natexlab#1{#1}\fi
\providecommand{\url}[1]{\texttt{#1}}
\providecommand{\href}[2]{#2}
\providecommand{\path}[1]{#1}
\providecommand{\DOIprefix}{doi:}
\providecommand{\ArXivprefix}{arXiv:}
\providecommand{\URLprefix}{URL: }
\providecommand{\Pubmedprefix}{pmid:}
\providecommand{\doi}[1]{\href{http://dx.doi.org/#1}{\path{#1}}}
\providecommand{\Pubmed}[1]{\href{pmid:#1}{\path{#1}}}
\providecommand{\bibinfo}[2]{#2}
\ifx\xfnm\relax \def\xfnm[#1]{\unskip,\space#1}\fi
\bibitem[{Czerwosz et~al.(2000{\natexlab{a}})Czerwosz, D{\l}u{\.z}ewski,
  Giera{\l}towski, Sobczak, Starnawska, and Wronka}]{n1}
\bibinfo{author}{E.~Czerwosz}, \bibinfo{author}{P.~D{\l}u{\.z}ewski},
  \bibinfo{author}{W.~Giera{\l}towski}, \bibinfo{author}{J.~Sobczak},
  \bibinfo{author}{E.~Starnawska}, \bibinfo{author}{H.~Wronka},
\newblock \bibinfo{title}{Electron emission from c/c+ pd films containing pd
  nanocrystals},
\newblock \bibinfo{journal}{Journal of Vacuum Science \& Technology B:
  Microelectronics and Nanometer Structures} \bibinfo{volume}{18}
  (\bibinfo{year}{2000}{\natexlab{a}}) \bibinfo{pages}{1064}.
\bibitem[{Czerwosz et~al.(2000{\natexlab{b}})Czerwosz, Dluzewski, Kozlowski,
  Sobczak, Starnawska, and Wronka}]{n2}
\bibinfo{author}{E.~Czerwosz}, \bibinfo{author}{P.~Dluzewski},
  \bibinfo{author}{M.~Kozlowski}, \bibinfo{author}{J.~Sobczak},
  \bibinfo{author}{E.~Starnawska}, \bibinfo{author}{H.~Wronka},
\newblock \bibinfo{title}{Electron emitting nanostructures of carbon+ pd
  system},
\newblock \bibinfo{journal}{Molecular Crystals and Liquid Crystals}
  \bibinfo{volume}{353} (\bibinfo{year}{2000}{\natexlab{b}})
  \bibinfo{pages}{237--242}.
\bibitem[{Czerwosz et~al.(2007)Czerwosz, Diduszko, D{\l}u{\.z}ewski,
  Keczkowska, Koz{\l}owski, Rymarczyk, and Sucha{\'n}ska}]{n3}
\bibinfo{author}{E.~Czerwosz}, \bibinfo{author}{R.~Diduszko},
  \bibinfo{author}{P.~D{\l}u{\.z}ewski}, \bibinfo{author}{J.~Keczkowska},
  \bibinfo{author}{M.~Koz{\l}owski}, \bibinfo{author}{J.~Rymarczyk},
  \bibinfo{author}{M.~Sucha{\'n}ska},
\newblock \bibinfo{title}{Properties of pd nanocrystals prepared by pvd
  method},
\newblock \bibinfo{journal}{Vacuum} \bibinfo{volume}{82} (\bibinfo{year}{2007})
  \bibinfo{pages}{372--376}.
\bibitem[{Russ(2011)}]{RUSS}
\bibinfo{author}{J.~C. Russ}, \bibinfo{title}{The image processing handbook},
  \bibinfo{publisher}{CRC press}, \bibinfo{year}{2011}.
\bibitem[{Waltz and Miller(1998)}]{WF}
\bibinfo{author}{F.~M. Waltz}, \bibinfo{author}{J.~W. Miller},
\newblock \bibinfo{title}{Efficient algorithm for gaussian blur using
  finite-state machines},
\newblock in: \bibinfo{booktitle}{Photonics East (ISAM, VVDC, IEMB)},
  \bibinfo{organization}{International Society for Optics and Photonics},
  \bibinfo{year}{1998}, pp. \bibinfo{pages}{334--341}.
\bibitem[{Gomez et~al.(2000)Gomez, Marroquin, and Sucar}]{Gomez}
\bibinfo{author}{G.~Gomez}, \bibinfo{author}{J.~Marroquin},
  \bibinfo{author}{L.~Sucar},
\newblock \bibinfo{title}{Probabilistic estimation of local scale},
\newblock in: \bibinfo{booktitle}{Pattern Recognition, 2000. Proceedings. 15th
  International Conference on}, volume~\bibinfo{volume}{3},
  \bibinfo{organization}{IEEE}, \bibinfo{year}{2000}, pp.
  \bibinfo{pages}{790--793}.
\bibitem[{Healey and Binford(1988)}]{Healey}
\bibinfo{author}{G.~Healey}, \bibinfo{author}{T.~O. Binford},
\newblock \bibinfo{title}{Local shape from specularity},
\newblock \bibinfo{journal}{Computer Vision, Graphics, and Image Processing}
  \bibinfo{volume}{42} (\bibinfo{year}{1988}) \bibinfo{pages}{62--86}.
\bibitem[{Hu and de~Haan(2006)}]{Gerard}
\bibinfo{author}{H.~Hu}, \bibinfo{author}{G.~de~Haan},
\newblock \bibinfo{title}{Low cost robust blur estimator},
\newblock in: \bibinfo{booktitle}{Image Processing, 2006 IEEE International
  Conference on}, \bibinfo{organization}{IEEE}, \bibinfo{year}{2006}, pp.
  \bibinfo{pages}{617--620}.
\bibitem[{Berg and Malik(2001)}]{Berg}
\bibinfo{author}{A.~C. Berg}, \bibinfo{author}{J.~Malik},
\newblock \bibinfo{title}{Geometric blur for template matching},
\newblock in: \bibinfo{booktitle}{Computer Vision and Pattern Recognition,
  2001. CVPR 2001. Proceedings of the 2001 IEEE Computer Society Conference
  on}, volume~\bibinfo{volume}{1}, \bibinfo{organization}{IEEE},
  \bibinfo{year}{2001}, pp. \bibinfo{pages}{I--607}.
\bibitem[{Geusebroek et~al.(2003)Geusebroek, Smeulders, and Van
  De~Weijer}]{Geusebroek}
\bibinfo{author}{J.-M. Geusebroek}, \bibinfo{author}{A.~W. Smeulders},
  \bibinfo{author}{J.~Van De~Weijer},
\newblock \bibinfo{title}{Fast anisotropic gauss filtering},
\newblock \bibinfo{journal}{Image Processing, IEEE Transactions on}
  \bibinfo{volume}{12} (\bibinfo{year}{2003}) \bibinfo{pages}{938--943}.
\bibitem[{PavloviC and Tekalp(1992)}]{Pavlovic}
\bibinfo{author}{G.~PavloviC}, \bibinfo{author}{A.~M. Tekalp},
\newblock \bibinfo{title}{Maximum likelihood parametric blur identification
  based on a continuous spatial domain model},
\newblock \bibinfo{journal}{Image Processing, IEEE Transactions on}
  \bibinfo{volume}{1} (\bibinfo{year}{1992}) \bibinfo{pages}{496--504}.
\bibitem[{Katsaggelos and Lay(1991)}]{Kgg}
\bibinfo{author}{A.~K. Katsaggelos}, \bibinfo{author}{K.-T. Lay},
\newblock \bibinfo{title}{Maximum likelihood blur identification and image
  restoration using the em algorithm},
\newblock \bibinfo{journal}{Signal Processing, IEEE Transactions on}
  \bibinfo{volume}{39} (\bibinfo{year}{1991}) \bibinfo{pages}{729--733}.
\bibitem[{Tabor and Spurek(2012)}]{SpuTab}
\bibinfo{author}{J.~Tabor}, \bibinfo{author}{P.~Spurek},
\newblock \bibinfo{title}{Cross-entropy clustering},
\newblock \bibinfo{journal}{Available from http://arxiv. org/pdf/1210.5594.
  pdf}  (\bibinfo{year}{2012}).
\bibitem[{{\'S}mieja and Tabor(2013)}]{CEC1}
\bibinfo{author}{M.~{\'S}mieja}, \bibinfo{author}{J.~Tabor},
\newblock \bibinfo{title}{Image segmentation with use of cross-entropy
  clustering},
\newblock in: \bibinfo{booktitle}{Proceedings of the 8th International
  Conference on Computer Recognition Systems CORES 2013},
  \bibinfo{organization}{Springer}, \bibinfo{year}{2013}, pp.
  \bibinfo{pages}{403--409}.
\bibitem[{Tabor and Misztal(2013)}]{CEC2}
\bibinfo{author}{J.~Tabor}, \bibinfo{author}{K.~Misztal},
\newblock \bibinfo{title}{Detection of elliptical shapes via cross-entropy
  clustering},
\newblock in: \bibinfo{booktitle}{Pattern Recognition and Image Analysis},
  \bibinfo{publisher}{Springer}, \bibinfo{year}{2013}, pp.
  \bibinfo{pages}{656--663}.
\bibitem[{Spurek et~al.(2013)Spurek, Tabor, and Zaj¹c}]{CEC3}
\bibinfo{author}{P.~Spurek}, \bibinfo{author}{J.~Tabor},
  \bibinfo{author}{E.~Zaj¹c},
\newblock \bibinfo{title}{Detection of disk-like particles in electron
  microscopy images},
\newblock in: \bibinfo{booktitle}{Proceedings of the 8th International
  Conference on Computer Recognition Systems CORES 2013},
  \bibinfo{organization}{Springer}, \bibinfo{year}{2013}, pp.
  \bibinfo{pages}{411--417}.
\bibitem[{Serra and Vincent(1992)}]{m_m_1}
\bibinfo{author}{J.~Serra}, \bibinfo{author}{L.~Vincent},
\newblock \bibinfo{title}{An overview of morphological filtering},
\newblock \bibinfo{journal}{Circuits, Systems and Signal Processing}
  \bibinfo{volume}{11} (\bibinfo{year}{1992}) \bibinfo{pages}{47--108}.
\bibitem[{Shih(2010)}]{Shih}
\bibinfo{author}{F.~Y. Shih}, \bibinfo{title}{Image processing and mathematical
  morphology: fundamentals and applications}, \bibinfo{publisher}{CRC press},
  \bibinfo{year}{2010}.
\bibitem[{Wilkinson and Roerdink(2009)}]{WR}
\bibinfo{author}{M.~H. Wilkinson}, \bibinfo{author}{J.~Roerdink},
\newblock \bibinfo{title}{Mathematical morphology and its application to signal
  and image processing},
\newblock in: \bibinfo{booktitle}{Proceedings of 9th International Symposium on
  Mathematical Morphology, Springer}, \bibinfo{year}{2009}.
\bibitem[{Di~Ruberto et~al.(2002)Di~Ruberto, Dempster, Khan, and
  Jarra}]{mor_aply_1}
\bibinfo{author}{C.~Di~Ruberto}, \bibinfo{author}{A.~Dempster},
  \bibinfo{author}{S.~Khan}, \bibinfo{author}{B.~Jarra},
\newblock \bibinfo{title}{Analysis of infected blood cell images using
  morphological operators},
\newblock \bibinfo{journal}{Image and Vision Computing} \bibinfo{volume}{20}
  (\bibinfo{year}{2002}) \bibinfo{pages}{133--146}.
\bibitem[{Nevatia(1976)}]{ad_mor_fill_2}
\bibinfo{author}{R.~Nevatia},
\newblock \bibinfo{title}{Locating object boundaries in textured environments},
\newblock \bibinfo{journal}{Computers, IEEE Transactions on}
  \bibinfo{volume}{100} (\bibinfo{year}{1976}) \bibinfo{pages}{1170--1175}.
\bibitem[{Nalwa and Pauchon(1987)}]{ad_mor_fill_3}
\bibinfo{author}{V.~S. Nalwa}, \bibinfo{author}{E.~Pauchon},
\newblock \bibinfo{title}{Edgel aggregation and edge description},
\newblock \bibinfo{journal}{Computer vision, graphics, and image processing}
  \bibinfo{volume}{40} (\bibinfo{year}{1987}) \bibinfo{pages}{79--94}.
\bibitem[{Mrozek et~al.(2012)Mrozek, {\.Z}elawski, Gryglewski, Han, and
  Krajniak}]{mrze}
\bibinfo{author}{M.~Mrozek}, \bibinfo{author}{M.~{\.Z}elawski},
  \bibinfo{author}{A.~Gryglewski}, \bibinfo{author}{S.~Han},
  \bibinfo{author}{A.~Krajniak},
\newblock \bibinfo{title}{Homological methods for extraction and analysis of
  linear features in multidimensional images},
\newblock \bibinfo{journal}{Pattern Recognition} \bibinfo{volume}{45}
  (\bibinfo{year}{2012}) \bibinfo{pages}{285--298}.
\bibitem[{Shih and Cheng(2004)}]{ad_mor_fill_1}
\bibinfo{author}{F.~Y. Shih}, \bibinfo{author}{S.~Cheng},
\newblock \bibinfo{title}{Adaptive mathematical morphology for edge linking},
\newblock \bibinfo{journal}{Information sciences} \bibinfo{volume}{167}
  (\bibinfo{year}{2004}) \bibinfo{pages}{9--21}.
\bibitem[{Hinz and Baumgartner(2003)}]{road_1}
\bibinfo{author}{S.~Hinz}, \bibinfo{author}{A.~Baumgartner},
\newblock \bibinfo{title}{Automatic extraction of urban road networks from
  multi-view aerial imagery},
\newblock \bibinfo{journal}{ISPRS Journal of Photogrammetry and Remote Sensing}
  \bibinfo{volume}{58} (\bibinfo{year}{2003}) \bibinfo{pages}{83--98}.
\bibitem[{Tupin et~al.(2002)Tupin, Houshmand, and Datcu}]{road_2}
\bibinfo{author}{F.~Tupin}, \bibinfo{author}{B.~Houshmand},
  \bibinfo{author}{M.~Datcu},
\newblock \bibinfo{title}{Road detection in dense urban areas using sar imagery
  and the usefulness of multiple views},
\newblock \bibinfo{journal}{Geoscience and Remote Sensing, IEEE Transactions
  on} \bibinfo{volume}{40} (\bibinfo{year}{2002}) \bibinfo{pages}{2405--2414}.
\bibitem[{Mahalanobis(1936)}]{Ma}
\bibinfo{author}{P.~C. Mahalanobis},
\newblock \bibinfo{title}{On the generalized distance in statistics},
\newblock \bibinfo{journal}{Proceedings of the National Institute of Sciences
  (Calcutta)} \bibinfo{volume}{2} (\bibinfo{year}{1936})
  \bibinfo{pages}{49--55}.
\bibitem[{Otsu(1975)}]{otsu}
\bibinfo{author}{N.~Otsu},
\newblock \bibinfo{title}{A threshold selection method from gray-level
  histograms},
\newblock \bibinfo{journal}{Automatica} \bibinfo{volume}{11}
  (\bibinfo{year}{1975}) \bibinfo{pages}{23--27}.
\bibitem[{Blum et~al.(1967)}]{skel1}
\bibinfo{author}{H.~Blum}, et~al.,
\newblock \bibinfo{title}{A transformation for extracting new descriptors of
  shape},
\newblock \bibinfo{journal}{Models for the perception of speech and visual
  form} \bibinfo{volume}{19} (\bibinfo{year}{1967}) \bibinfo{pages}{362--380}.
\bibitem[{Attali et~al.(1995)Attali, di~Baja, and Thiel}]{pruning1}
\bibinfo{author}{D.~Attali}, \bibinfo{author}{G.~S. di~Baja},
  \bibinfo{author}{E.~Thiel},
\newblock \bibinfo{title}{Pruning discrete and semicontinuous skeletons},
\newblock in: \bibinfo{booktitle}{Image Analysis and Processing},
  \bibinfo{organization}{Springer}, \bibinfo{year}{1995}, pp.
  \bibinfo{pages}{488--493}.
\bibitem[{Shaked and Bruckstein(1998)}]{pruning2}
\bibinfo{author}{D.~Shaked}, \bibinfo{author}{A.~M. Bruckstein},
\newblock \bibinfo{title}{Pruning medial axes},
\newblock \bibinfo{journal}{Computer vision and image understanding}
  \bibinfo{volume}{69} (\bibinfo{year}{1998}) \bibinfo{pages}{156--169}.
\bibitem[{Lam et~al.(1992)Lam, Lee, and Suen}]{thin1}
\bibinfo{author}{L.~Lam}, \bibinfo{author}{S.-W. Lee}, \bibinfo{author}{C.~Y.
  Suen},
\newblock \bibinfo{title}{Thinning methodologies-a comprehensive survey},
\newblock \bibinfo{journal}{IEEE Transactions on pattern analysis and machine
  intelligence} \bibinfo{volume}{14} (\bibinfo{year}{1992})
  \bibinfo{pages}{869--885}.
\bibitem[{ima(????)}]{imageJ}
\bibinfo{title}{Local gaussian filter and adaptive morphology operation as a
  plug--in fro imagej},
  \bibinfo{howpublished}{\url{http://ww2.ii.uj.edu.pl/~spurek/imageJ/LocalGaussianFilter_AdaptiveMorphologyOperation/LocalGaussianFilter_AdaptiveMorphologyOperation.html}},
  \bibinfo{note}{Accessed: 2012-12-10}.

\end{thebibliography}
\end{document}